\documentclass{article}

\usepackage{fontenc}
\usepackage{amsmath}
\usepackage{graphicx}
\usepackage{subfigure}
\usepackage{color}

\begin{document}

\title{\Large \bf CHAOTIC DYNAMICS OF FOREST FIRES}

\author{
K.~Malarz$^*$,
S.~Kaczanowska,
K.~Ku{\l}akowski$^\dag$\\
\normalsize \em
Department of Theoretical and Computational Physics,
Faculty of Physics and Nuclear\\
\normalsize \em
Techniques,
University of Mining and Metallurgy (AGH)\\
\normalsize \em
al. Mickiewicza 30, PL-30059 Krak\'ow, Poland.\\
\normalsize \em
E-mail: $^*$malarz@agh.edu.pl, $^\dag$kulakowski@novell.ftj.agh.edu.pl
}

\maketitle

\begin{abstract}
\noindent
In the thermodynamic limit, a probabilistic cellular automaton can be
approximated by a deterministic nonlinear map. Here we construct such a map
for the forest fire problem. The construction is based on the results of the
Monte Carlo simulation, performed on a square lattice of million cells. The
results of the calculation are analyzed by means of the Hoshen--Kopelman
algorithm (HKA). The only parameter of the map describes the probability that
a tree appears at an empty cell during one time step. The obtained map seems
to be non-differentiable at the percolation threshold. The Lyapunov exponent
for the map is positive. Also, we found the cycle of length three by means of the
method of symbolic dynamics. The results are illustrated by the experimental
data on the forest fires in Canada in years 1970--2000. Although these data are
fortunately far from thermodynamic limit, their qualitative character is
reproduced for smaller lattices.
\end{abstract}


\section{Introduction}
The problem of forest fires has got a wide audience \cite{duarte} for its
percolation origin \cite{stauffer} and rich connections to basic problems
in computational physics, e.g. to the self-organized criticality. On the
other
hand, the problem is not entirely virtual, although experiments are not
eagerly accepted by the society. The data can be found in annual statistics
of forest rangers and insurance companies. Both institutions are obviously
interested in predicting of forest fires, at least in the statistical sense.
However, there are some indications that such predictions are not possible
in long time scale | the problem is chaotic \cite{chenbak,socolar,malarz}.

The present work is a continuation of Ref. \cite{malarz}. In that paper,
a map $p_n\to p_{n+1}$ which describes the forest density $p$ in subsequent
time steps $n$ was proposed. The construction was based on the idea of
reducing
a probabilistic cellular automaton to the map, which is equivalent to the
mean
field approximation of  the automaton rules \cite{wooters}. The map was
designed not so as to reproduce the dynamics of the fire front, but rather to
simulate the statistics of the forest density in subsequent years $n$. The
forest was represented by a square lattice $L\times L$, occupied by trees
with
probability $p$. The above map was equivalent to a superposition of two
functions, where each function transforms the forest concentration $p$. The
growing process was represented by a parabolic function,
$p_{n+1}=p_n+rp_n(1-p_n)$, where $r$ is a parameter. Next, a season of fires
was modeled by a reduction of $p_{n+1}$ by what was burned. This amount was
demonstrated to be equal to the mean squared size of clusters of trees. By
definition, a tree belonged to a cluster if it was a nearest neighbor of
another tree which belongs to it. The neighborhood was of von Neumann kind.

The prescription of calculating the burned area contained the
probability that a cluster is ignited. This probability can be defined ``per
area'' or ``per cluster''. We have obtained two kinds of statistics, which can
be
interpreted as follows: In the former case, a lightning strikes at a cell,
occupied by a tree or not. The probability that the ignited tree belongs to a
cluster is proportional to the size of this cluster, i.e. the number of trees
in it. The burned area is also proportional to this size. Then, the totally
burned area is the mean squared size of the cluster, as stated above. The
weight of the cluster of size $s_i$ is $w_i'=s_i/L^2$. In the latter case, the
fire is set intentionally, what means that a tree is ignited with probability
one. The rest of argumentation is the same as in the former case. The
appropriate weight in a such situation is
\begin{equation}
\label{eq_1}
w_i''=s_i/\sum_i s_i.
\end{equation}
The former case ($w'$) was discussed thoroughly in Ref.~\cite{malarz}. In the
present
work we focus on the latter case, where the weights are given by Eq. \eqref{eq_1}.

The paper is organized as follows: The next section is devoted to the
calculation of the map $p_n\to p_{n+1}$, which is obtained by an application
of the HKA \cite{hoshen}. In Section 3 we demonstrate
the results on the Lyapunov exponent and the symbolic dynamics, which allows
to
obtain the cycle of the length three in the time evolution of the trees density
$p$.
There, we also refer to the experimental data. Our conclusions are given in
the last section.

\section{The model}
The map $p_n \to p_{n+1}$ is a superposition of two functions.
The first one, $p \to p+rp(1-p)$, describes the season of growing.
As a result, trees appear in empty cells with the probability $rp(1-p)$, where $r$ is a parameter from the range $[0,1]$.
Afterwards, the burned area of the forest is calculated with the weights $w_i''$, i.e.
\begin{equation}
\label{eq_2}
A(p)=\frac{1}{L^2}\sum_i w_i'' s_i
\end{equation}
and it is subtracted from trees concentration after growth season, i.e. $p \to p-A(p)$.

\begin{figure}
\begin{center}
\includegraphics[angle=-90,width=0.9\textwidth]{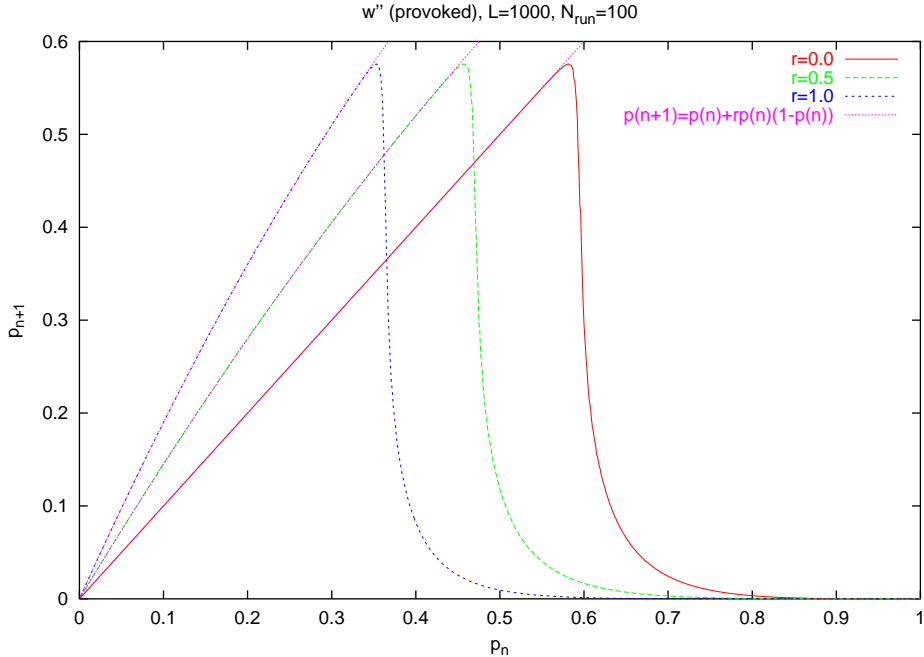}
\end{center}
\caption{The map for provoked fires ($w''$) and different values of the parameter $r$.}
\label{fig_map}
\end{figure}

The cluster size distribution is obtained by means of HKA.
With HKA \cite{hoshen} we are able to label all
occupied sites on a $L\times L$ large square lattice in such a way that the
sites with the same label belong to the same cluster and different labels are
associated to different clusters.
As a result, we obtain the cluster size distribution for $1001$ values of $p$, $p_i=10^{-3}i$, $i=0,1,2,...,10^3$.
The simulation is carried out on a square lattice $10^3\times 10^3$ large, with a fraction of $p+rp(1-p)$ occupied sites for different values of the initial trees concentration $0\le p\le 1$.
The map is got for 1001 equidistant values of $p$ as an average over hundred
trial lattice configurations for each value of $p_i$.
For other values of $p$ we use straight lines joining neighboring points $p_i$ and $p_{i+1}$.
The obtained map is shown in Fig. \ref{fig_map} for three different values of the parameter $r$.

The method above suffers from the finite size of the lattice and the finite grid.
In particular, we observe some shift of the maximum of the map for $r=0$ with respect to the percolation threshold for the square lattice.
Instead to occur at $p_c=0.59273$ \cite{stauffer}, we find the maximum near $p=0.583$.
The rounding effect of the finite size reduces the map to the well-known unimodal map, where the period-doubling bifurcation must occur, at least at some range of the parameter $r$.
That is why we extrapolate the shape of the curve to a tentative limit of infinite size (the so-called thermodynamic limit).
The result for $r=0$ is
\begin{equation}
\label{eq_3}
p_{n+1}=
\begin{cases}
p_n & \iff p_n\le p_c,\\
p_c \exp(-12.735\sqrt{p_n-p_c} ) & \iff p_n>p_c.
\end{cases}
\end{equation}

\section{Results}
The symbolic dynamics is known to be the most rigorous method to investigate
chaos in one-dimensional maps \cite{haobai}.
The basic tool is
to assign symbolic words to
the periodic cycles.
Example giving, the cycle of length two is represented by CR, of length three
| by CRL, etc.
The appearance of the cycle three in the time evolution of a given variable is known to be a proof of
the presence of cycles of all possible lengths.
In our case the cycles of the variable $p$ are not superstable, because the curve given by Eq. \eqref{eq_3} is not differentiable at $p=p_c$.
In fact, the cycles are even not stable: we do not observe any windows of stability in the bifurcation diagram.
Still, it is not difficult to find a cycle three by the method of ``shooting''. Fixing the value of $p_1$, we change the parameter $r$ until $p_4=p_1$. An example is the cycle $p=0.340000, 0.419593, 0.505974$ for $r=0.354694$.

\begin{figure}
\begin{center}
\subfigure{
\label{fig_lya_a}
(a) \includegraphics[angle=-90,width=0.9\textwidth]{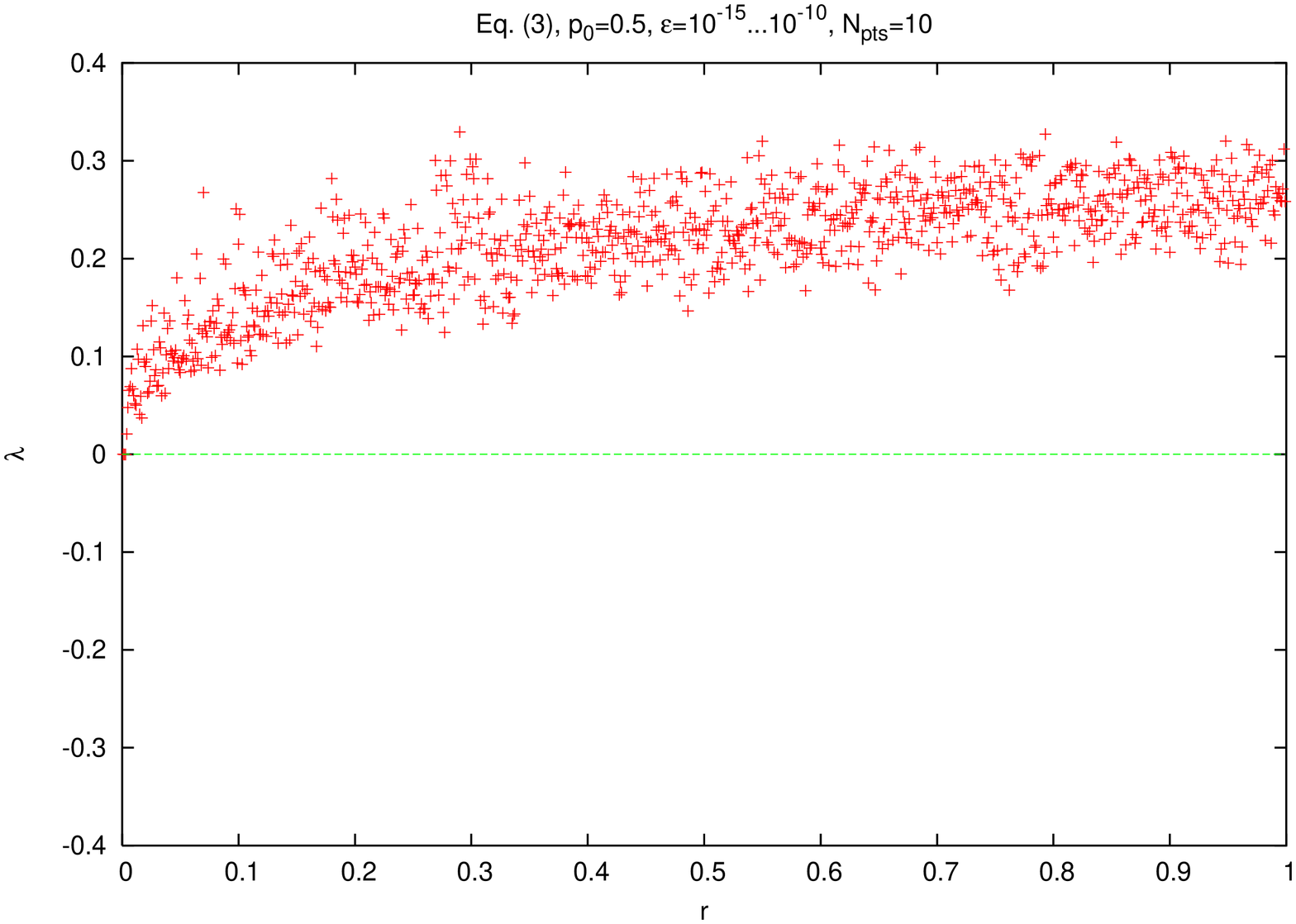}
}
\subfigure{
\label{fig_lya_b}
(b) \includegraphics[angle=-90,width=0.9\textwidth]{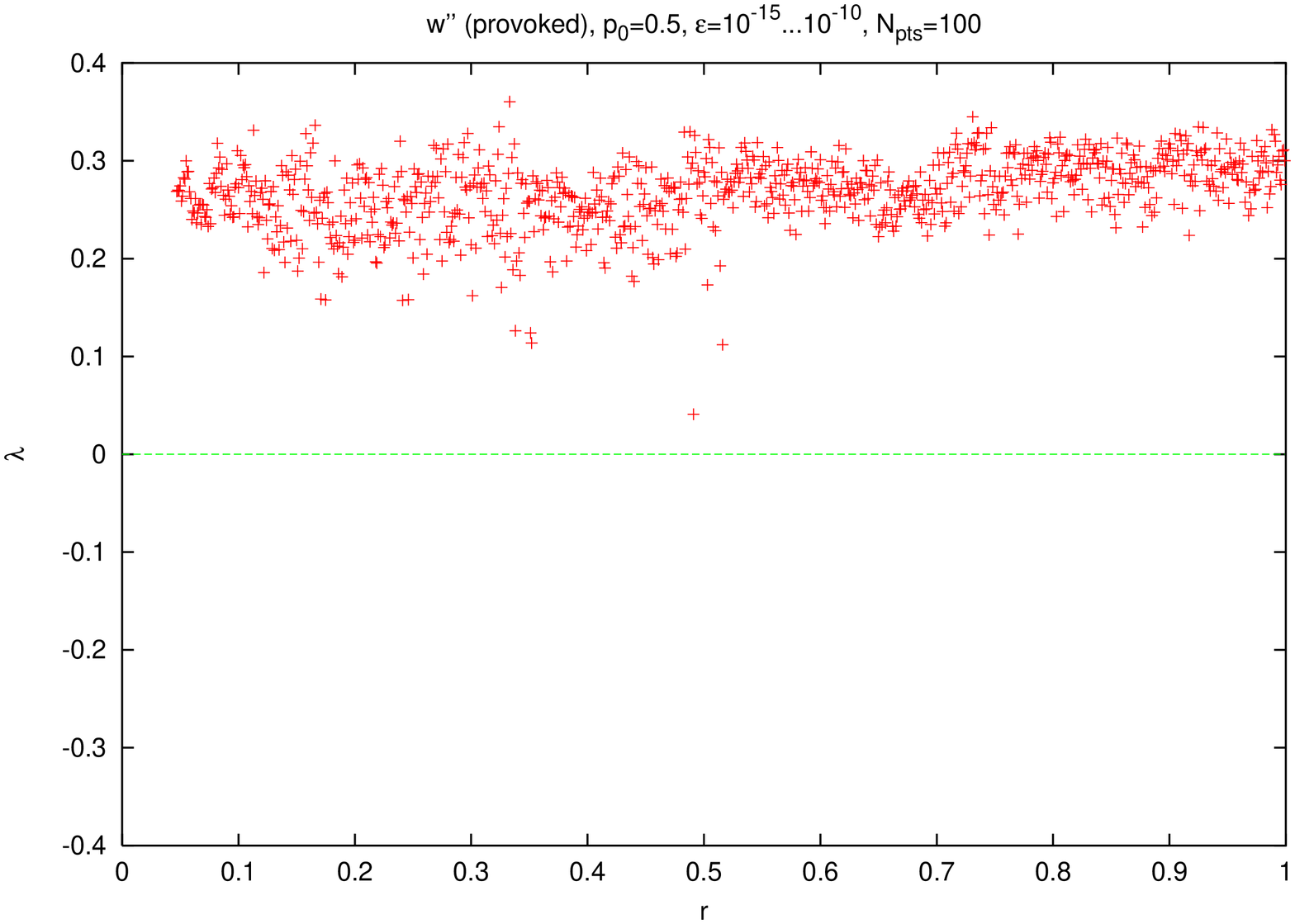}
}
\end{center}
\caption{Linear fit to the first fifty points of Fig. \ref{fig_cyc} allows
to evaluate the Lyapunov exponent $\lambda$ for different $r$ and (a) infinitely
large or (b) finite lattice.
Only positive values of $\lambda$ are presented.}
\label{fig_lya}
\end{figure}

\begin{figure}
\begin{center}
\includegraphics[angle=-90,width=0.9\textwidth]{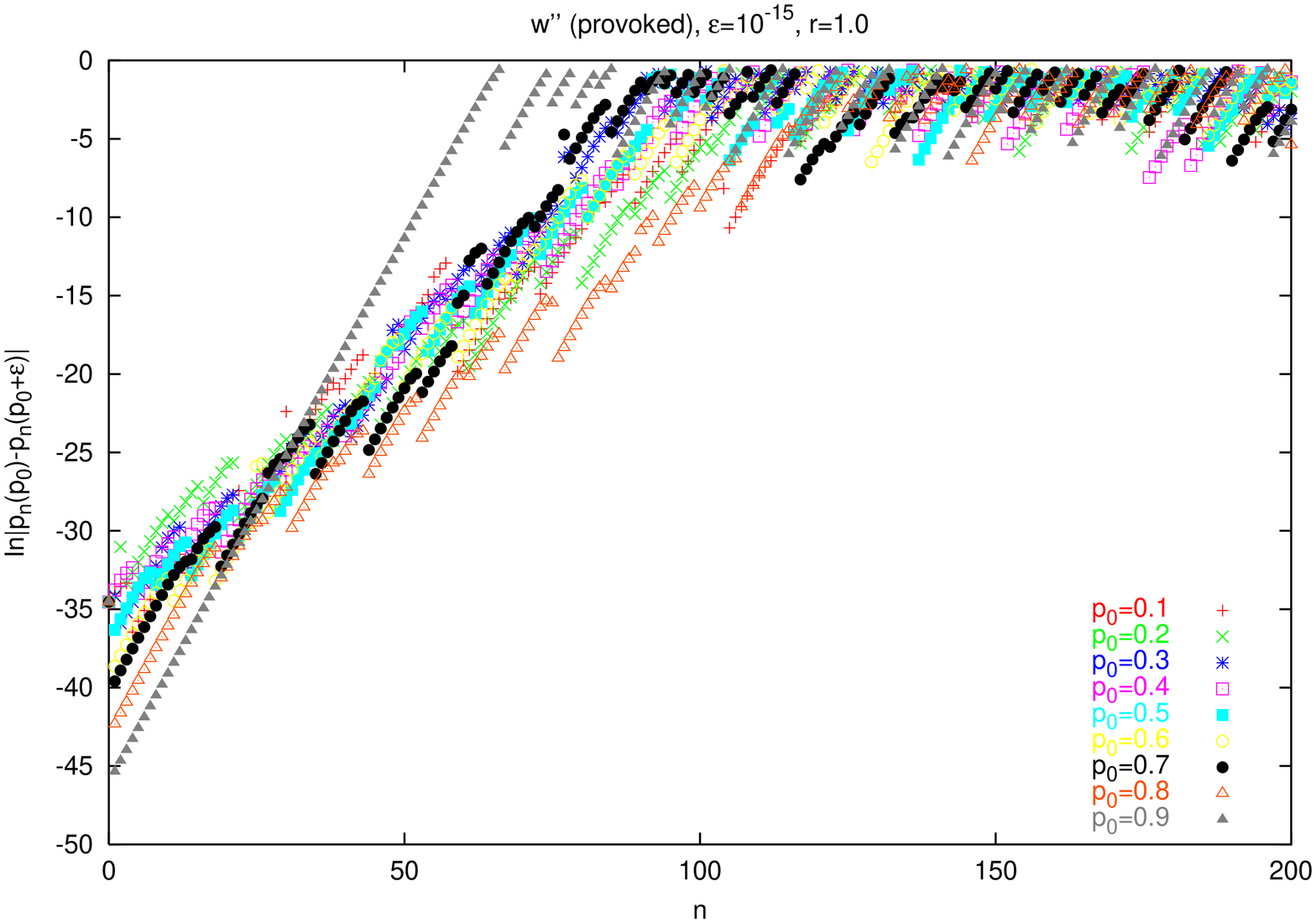}
\end{center}
\caption{Subsequent differences between two trajectories evolution which initially
differ on $\varepsilon$ and different initial concentration of trees $p_0$.
Here $r=1$ and $\varepsilon=10^{-15}$.}
\label{fig_cyc}
\end{figure}

In Fig. \ref{fig_lya} we show the values of the Lyapunov exponent as dependent on the parameter $r$.
These results are obtained for the extrapolated map given by the Eq. \eqref{eq_3} and by the method described in Ref. \cite{bennetin} (Fig. \ref{fig_lya_a}).
To obtain the Lyapunov exponent we have investigated the subsequent
differences between two initially nearby trajectories.
The slope of the time dependence of the logarithm of this difference (see
Fig. \ref{fig_cyc}) gives the exponent $\lambda$.
The results for the finite lattice are qualitatively the same (Fig. \ref{fig_lya_b}).

\begin{figure}
\begin{center}
\subfigure{
\label{fig_exp_1a}
(a) \includegraphics[angle=-90,width=0.9\textwidth]{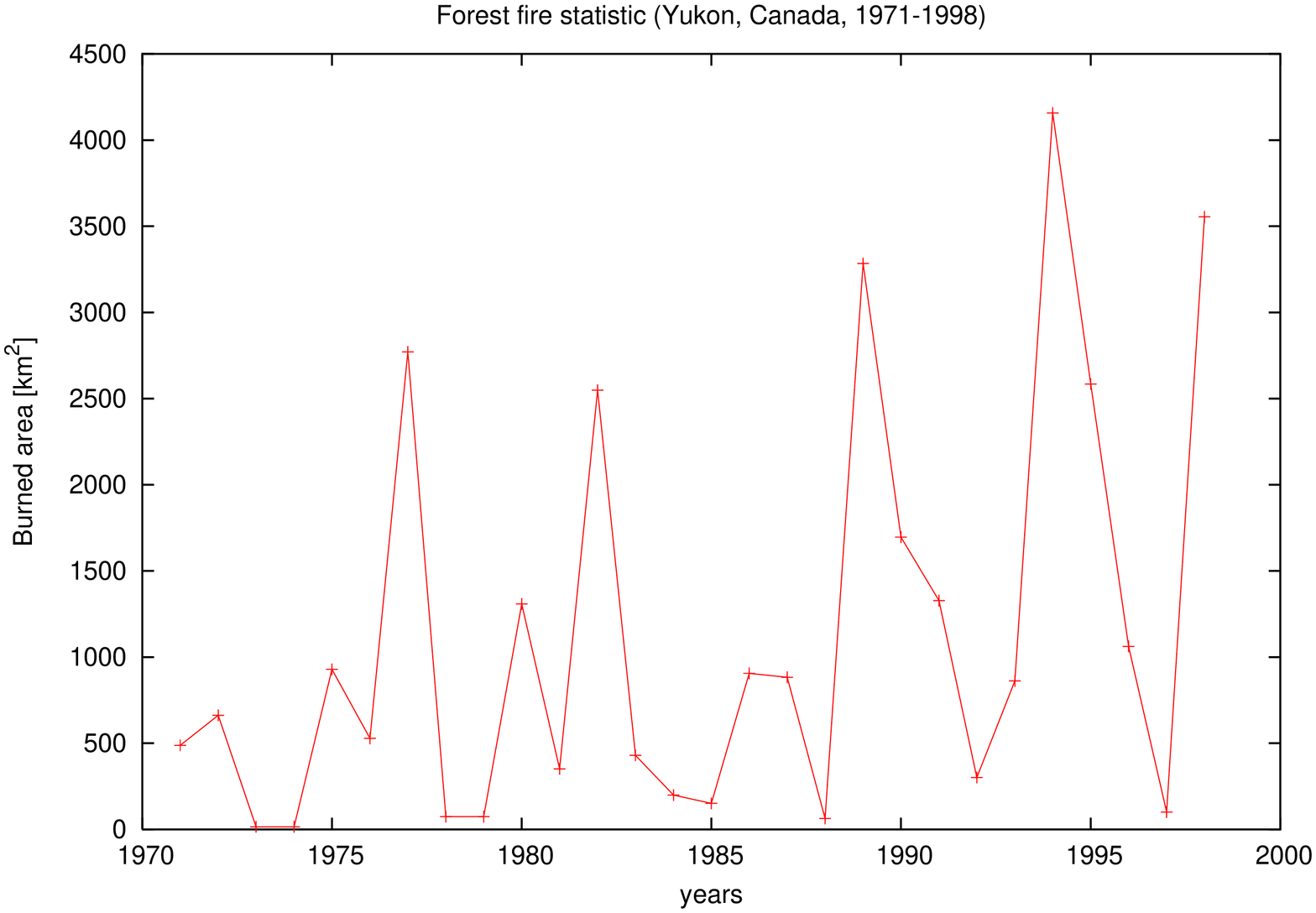}
}
\subfigure{
\label{fig_exp_1b}
(b) \includegraphics[angle=-90,width=0.9\textwidth]{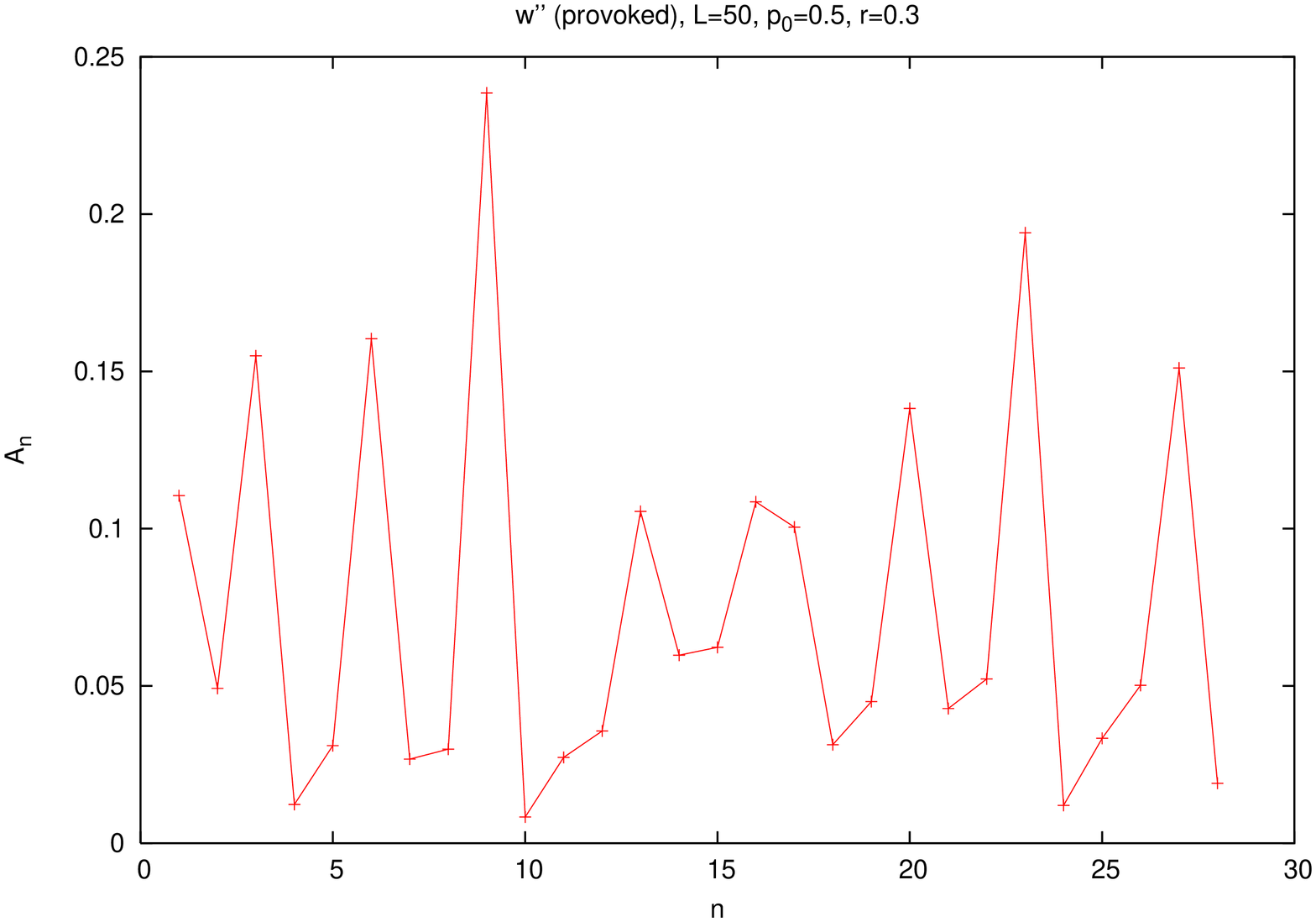}
}
\end{center}
\caption{The comparison of the forest fire statistics: Time series of fire
sizes (a) for Yukon, Canada, 1971--1998 and (b) the computer simulation on small lattice.}
\label{fig_exp_1}
\end{figure}

\begin{figure}
\begin{center}
\subfigure{
\label{fig_exp_2a}
(a) \includegraphics[angle=-90,width=0.9\textwidth]{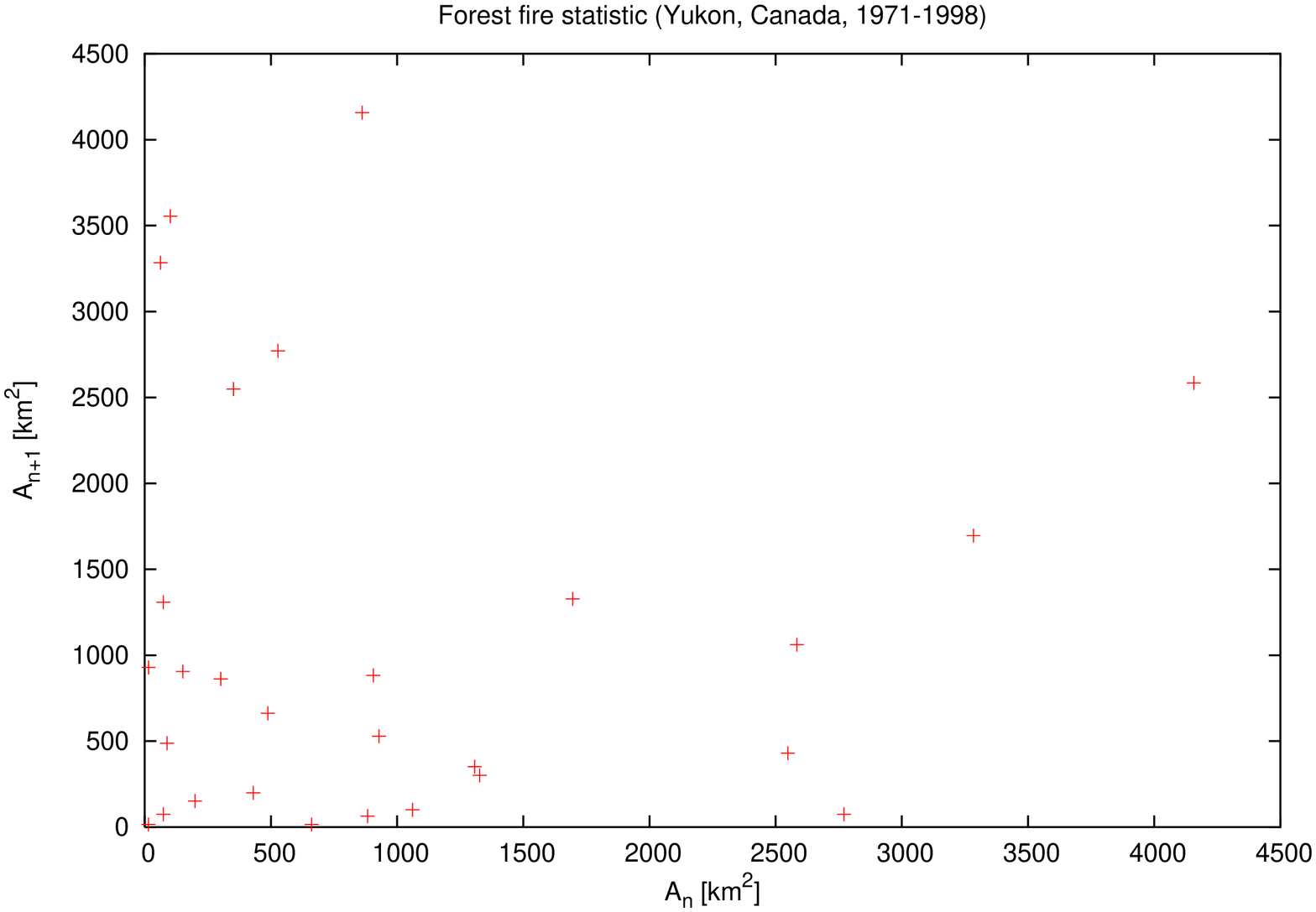}
}
\subfigure{
\label{fig_exp_2b}
(b) \includegraphics[angle=-90,width=0.9\textwidth]{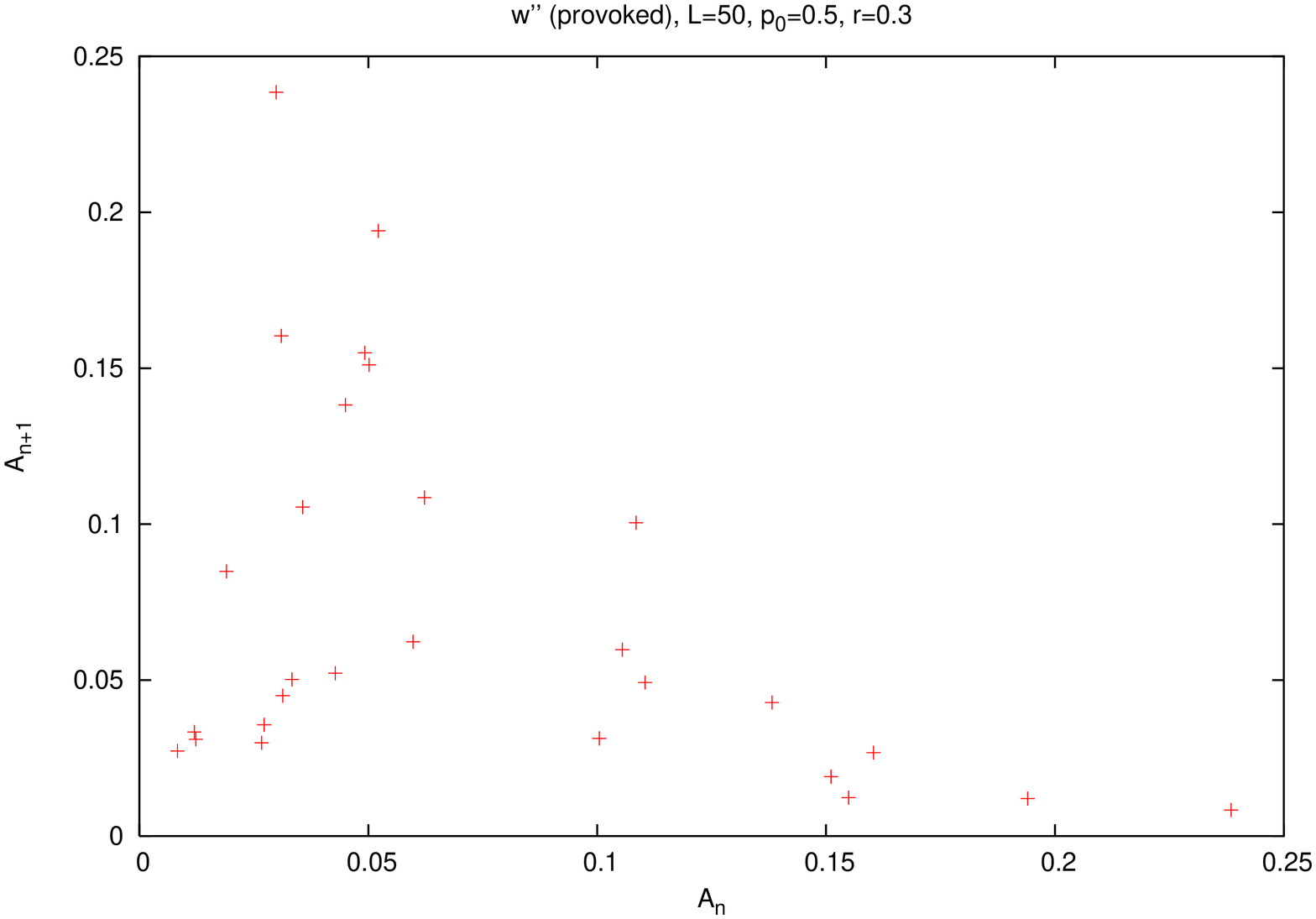}
}
\end{center}
\caption{The comparison of the forest fire statistics: $A(p_{n+1})$ vs.
$A(p_n)$ (a) for Yukon, Canada, 1971--1998 and (b) the computer simulation on small lattice.}
\label{fig_exp_2}
\end{figure}

In Fig. \ref{fig_exp_1} we show two plots: calculated time dependence of
$A(p)$ for the lattice of $50\times 50$ trees (Fig. \ref{fig_exp_1b}) and the
statistical data on the forest fires in Yukon \cite{malarz,web} (Fig. \ref{fig_exp_1a}).
We have also checked that the phase portraits $A(p_{n+1})$ vs. $A(p_n)$ are more or less the same as the appropriate phase portraits for the experimental data \cite{web} (Fig. \ref{fig_exp_2}).
As we can see, small fires occur more often than large ones.
Both kinds of the phase portraits show some anticorrelations between the forest fires in subsequent years: large burned area in a given year is likely to be followed by a small area in the subsequent year.
This remains true also for the data on the forest fires in other provinces of Canada in years 1970--2000.

\begin{figure}
\begin{center}
\includegraphics[angle=-90,width=0.9\textwidth]{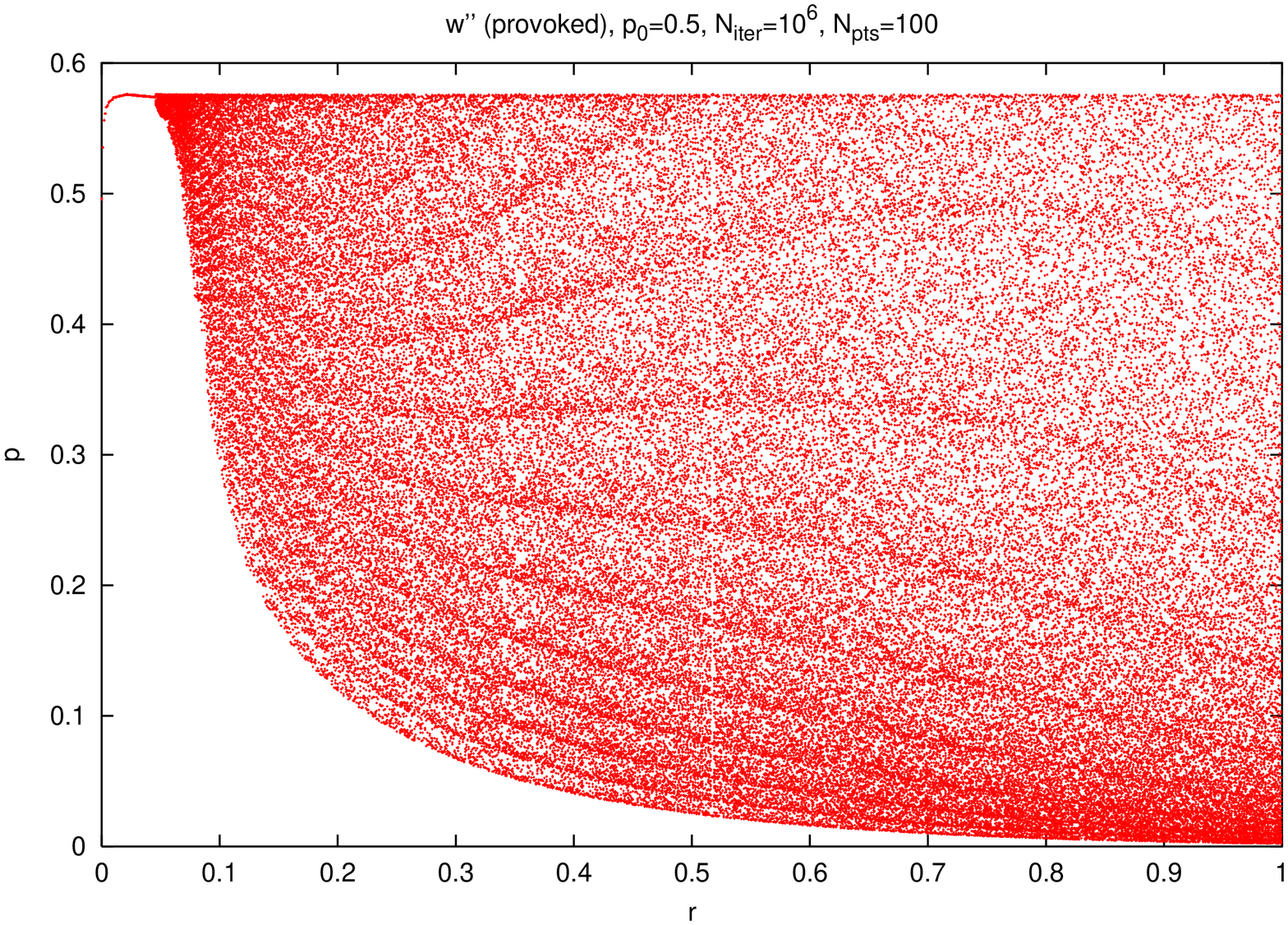}
\end{center}
\caption{The bifurcation diagram for the map obtained with HKA.}
\label{fig_bif}
\end{figure}

\begin{figure}
\begin{center}
\includegraphics[angle=-90,width=0.9\textwidth]{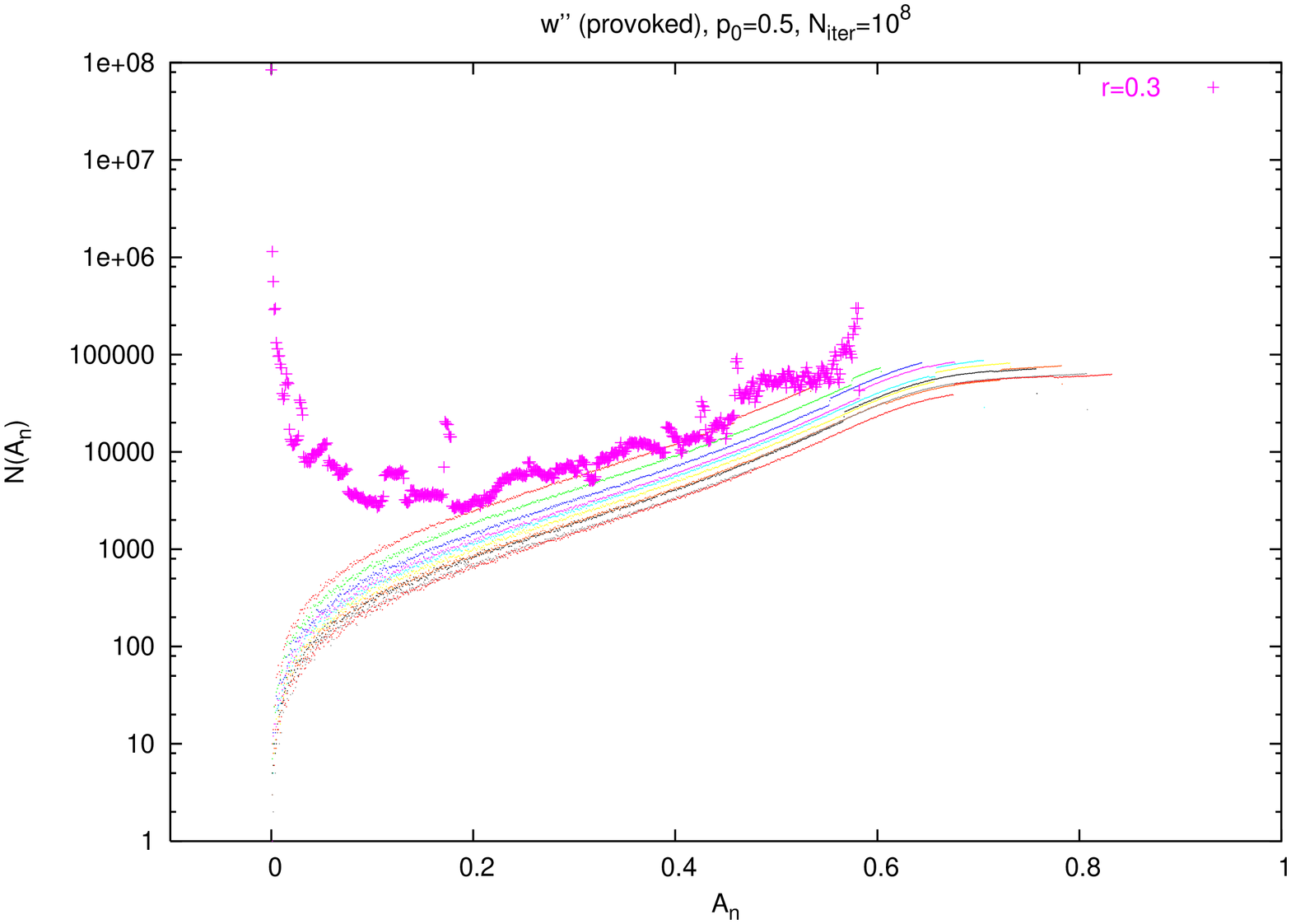}
\end{center}
\caption{The histogram of burned area obtained for the map obtained by HKA (pluses) and by means of computer simulation for the map given by Eq. \eqref{eq_3} (dots).
Subsequent lines of dots correspond to larger and larger value of parameter
$r=0.1,0.2,\cdots,0.9$ from the picture top to bottom.} 
\label{fig_hist}
\end{figure}

The bifurcation diagram for the map obtained by HKA (shown in Fig. \ref{fig_map}) presents itself as a homogeneous spot (see Fig. \ref{fig_bif}).
The same is true for the bifurcation diagram for the analytic map defined by Eq. \eqref{eq_3}.
The histogram of the forest fires for the map obtained by HKA is presented in Fig. \ref{fig_hist}, together with the histogram for the analytic map.
The difference is that in the latter case, small fires do not occur.
Still however, the events without fires are the most frequent in all the
cases.
Both histograms show large amount of events without fires, and the exponential shape in the range of intermediate and large fires.
In this region $N(A)$ increases with $A$, on the contrary to the predictions of the models which display the self-organized criticality \cite{drossel}. This can be
interpreted as a consequence of our assumption, that large clusters are
ignited more likely. The obtained curve is shifted to right when the
parameter
$r$ increases, as it is shown in Fig. \ref{fig_hist} for the analytic map.

\section{Conclusions}
The results presented allow to state that it is not possible to predict the
statistics of the forest fires in long time scale.  This conclusion is drawn
from our much simplified model of the fires, where the forest is reduced to a
lattice of cells, and a tree ignition | to the nearest-neighbor relation
with
a tree which already burns. Obviously, almost all technical details of a real
fire \cite{duarte} are omitted in this approach. Still we believe that our
conclusion is justified. The argument is as follows: suppose that a much
simplified model predicts chaos. Suppose that the model has been improved by
introducing several complex details. Is it possible that the new version
gives
a regular and predictable behavior? The answer is: no. The chaotic character
could be eliminated by a noise, but the predictability cannot be improved
\cite{warning}.

\section*{Acknowledgements}
The authors thank to Professor Dietrich Stauffer for valuable comments.
The simulations were carried out in ACK-CYFRONET-AGH. The
time on SGI 2800 machine is financed by the Polish Committee for Scientific
Research (KBN) with grant No. KBN/\-SGI2800/\-022/\-2002.


\end{document}